\documentclass[journal]{IEEEtran}

\pdfoutput=1
\usepackage{ifpdf}
\usepackage{cite}
\usepackage[pdftex]{graphicx}
\usepackage{amsmath}
\usepackage{amsfonts}
\usepackage{array}
\usepackage{float}
\usepackage{fancyhdr}
\usepackage[colorlinks,linkcolor=black,anchorcolor=black,citecolor=black]{hyperref}
\usepackage{epstopdf}

\usepackage{multirow}

\usepackage{algorithm}
\usepackage{algpseudocode}

\usepackage{booktabs}
\usepackage[caption=false,font=footnotesize]{subfig}

\usepackage{paralist}
\let\itemize\compactitem
\let\enditemize\endcompactitem

\usepackage{caption3}

\begin{document}



\title{Decentralized Baseband Processing with   Gaussian Message Passing  Detection for    Uplink  Massive MU-MIMO Systems}


\author{Zhenyu~Zhang, Yuanyuan~Dong,   Keping~Long,~\IEEEmembership{Senior~Member,~IEEE}, Xiyuan~Wang,~\IEEEmembership{Member,~IEEE}, and Xiaoming~Dai,~\IEEEmembership{Member,~IEEE}
\thanks{This work was supported by the National Natural Science Foundation of China under Grant 61871029. (\emph{Corresponding author: Xiaoming Dai.})}
\thanks{Z. Zhang, Y. Dong, K. Long,  and X. Dai are with the School of Computer and Communication Engineering, University of Science and Technology Beijing, Beijing 100083, China (e-mail: zzydq1@163.com; daixiaoming@ustb.edu.cn).}
\thanks{X. Wang is with  Beijing Information Science and Technology University, Beijing 100101, China.}
}

\maketitle


\begin{abstract}

Decentralized baseband processing (DBP) architecture, which partitions the base station antennas into multiple antenna clusters, has been recently proposed to alleviate the excessively high interconnect bandwidth, chip input/output data rates, and detection complexity for massive multi-user multiple-input multiple-output (MU-MIMO) systems. In this paper, we develop a novel decentralized Gaussian message passing (GMP) detection for the DBP architecture. By projecting the discrete probability distribution into a complex Gaussian function, the local means and variances iteratively calculated in each antenna cluster are fused to generate the global symbol beliefs based on the proposed message fusion rule in the centralized processing unit. We present the framework and analysis of the convergence of the decentralized GMP detection based on state evolution under the assumptions of large-system limit and Gaussian sources. Analytical results corroborated by simulations demonstrate that nonuniform antenna cluster partition scheme exhibits higher convergence rate than the uniform counterpart. Simulation results illustrate that the proposed decentralized GMP detection outperforms the recently proposed decentralized algorithms.

\end{abstract}



\IEEEpeerreviewmaketitle


\section{Introduction}
\label{sectI}


\IEEEPARstart{M}{assive}   multi-user  multiple-input multiple-output (MU-MIMO)  systems, in which hundreds  of antennas  are equipped at the base station (BS),
have been extensively investigated  owing to the large gains in  spectral efficiency, capacity, and  reliability  over  traditional 
small-scale MIMO systems \cite{m-mimo-1}.
However, one of the  most critical implementation challenges is the  excessively high amount of raw baseband data  that must be  transferred between the BS radio frequency (RF)  units  and the baseband
processing unit \cite{dbp-1,dbp-2}.
For instance, the raw baseband data rates exceed $200$ Gbit/s for a massive MU-MIMO  BS operating at $40\thinspace \text{MHz}$ bandwidth with $128$ BS antennas and $10$-bit analog-to-digital   converters \cite{dbp-2,dbp-3}.
Such high  data rates exceed  the bandwidth of existing interconnect technologies and approach the limit of  existing chip input/output  interfaces \cite{dbp-4,dbp-5}.
Furthermore, classical  detection algorithms typically rely on centralized baseband processing.
This centralized framework requires  full channel state information (CSI) and full received signal, which
brings excessively  high computational  complexity and power consumption for massive MU-MIMO systems \cite{dbp-2}.

To mitigate this challenge,
decentralized baseband processing (DBP) \cite{dbp-1,dbp-2,dbp-3,dbp-4,dbp-5} architecture has been recently  proposed for massive MU-MIMO systems.
In this decentralized architecture, the BS  antennas are partitioned  into multiple antenna clusters with independent RF  circuitry and computing hardware.
Each antenna cluster performs decentralized channel estimation (CHEST) and  signal detection, i.e., only the  local CSI and received signal are required in each antenna cluster.
A centralized processing unit is followed to generate the global estimated symbols based on a given fusion rule for  decoding.
Reference  \cite{dbp-1} detailed the decentralized   maximum ratio combining (MRC) and   minimum mean square error (MMSE) detections, and proposed an optimal fusion rule utilizing the weighted average of local estimates. The MMSE  algorithm  involves complicated matrix inversion whose computational complexity is cubic  to the user  number, which is unfavorable in practical implementation.
Matrix inversion-less   decentralized conjugate gradient (CG) \cite{dbp-2,dbp-3}, alternating direction method of multipliers (ADMM) \cite{dbp-2}, and coordinate descent (CD) \cite{dbp-4} signal detections have been   proposed to reduce the complexity of MMSE detection.
However, the  bit error rate (BER) performance  of these detections  only approaches the MMSE method.
The authors in \cite{dbp-1,dbp-5} proposed a nonlinear detection scheme that builds upon the large-MIMO approximate message passing (LAMA) algorithm, which achieves a slight performance gain over the decentralized MMSE.
Note that all these DBP-based   detections  consider the uniform  antenna cluster partition, in which  the BS antennas are partitioned equally.

In this paper, we propose an efficient DBP-based detection scheme
under the framework of the Gaussian message passing (GMP) algorithm \cite{gmp-1,gmp-2,gmp-3} for uplink massive MU-MIMO systems.
The GMP algorithm, which is operated on a fully-connected loopy factor graph \cite{gmp-4}, has been extensively studied for signal detection   in massive MU-MIMO systems.
In the proposed decentralized GMP detection, each antenna cluster executes independent CHEST and GMP detection in parallel  and propagates the local messages to the centralized processing unit.
A  novel fusion rule is proposed based on the message passing rule to form the global symbol beliefs and estimated symbols, rather than using the weighted average scheme proposed in \cite{dbp-1,dbp-4}.
To prove the convergence of the proposed GMP algorithm, the state evolution  is adopted to track the variance variation under the assumptions of large-system limit and Gaussian sources.
Furthermore, we analyze the antenna cluster partition scheme and demonstrate that
the  nonuniform partition results in a smaller symbol belief variance and higher convergence compared with the uniform partition when  fixing the antenna cluster number.
Numerical results  illustrate that the   nonuniform antenna cluster partition scheme  achieves performance gain over the   uniform counterpart.
In addition, the proposed decentralized GMP detection outperforms the recently proposed  decentralized algorithms and exhibits linear computational complexity.

The remainder of this paper is organized as follows.
Section \ref{sectII} describes the system model of uplink  massive MU-MIMO with DBP architecture.
In Section \ref{sectIII}, we  present the proposed decentralized GMP detection.
Simulation results are illustrated  in Section \ref{sectIV}.
Finally, conclusions are drawn in Section \ref{sectV}.

\emph{Notation}:
Boldface uppercase letter $\mathbf{X}$ and lowercase letter $\mathbf{x}$ denote matrices and column vectors, respectively.
${{\left( . \right)}^{T}}$, ${{\left( . \right)}^{H}}$,  ${{\left( . \right)}^{-1}}$, and ${{\left( . \right)}^{*}}$ denote the transpose, conjugate transpose, matrix inversion, and complex conjugate, respectively.
$\mathbb N$ is the set of positive integers.
${{\mathbf{I}}_{N}}$ represents the identity matrix with dimension $N$.
${{x}_{k}}$ represents the $k\text{th}$ element of $\mathbf{x}$.
The $n\text{th}$ row and $k\text{th}$ column element of $\mathbf{X}$ is denoted by ${{x}_{n,k}}$.
$\mathbb{E}\left[ . \right]$ represents the expectation operator.
The  complex Gaussian probability distribution function (PDF)
is denoted by
$\mathcal{C}\mathcal{N}\left( x;\mu ,{{\sigma }^{2}} \right) ={{\left( \pi {{\sigma }^{2}} \right)}^{-1}}\exp \left( -{{{\left| x-\mu  \right|}^{2}}}/{{{\sigma }^{2}}}\; \right)$
with mean $\mu $ and variance ${{\sigma }^{2}}$.

\section{System Model}
\label{sectII}

\subsection{Uplink Massive MU-MIMO Systems}

Consider an uplink massive MU-MIMO system with $K$ single-antenna  users  simultaneously transmitting to  a BS with $N$ $(N \gg K)$ antennas.
The  complex baseband input-output relation of the uplink massive MU-MIMO channel is given by
\begin{equation}
\label{equ1}
\mathbf{y}=\mathbf{Hx}+\mathbf{n},
\end{equation}
where  $\mathbf{y}\in {{\mathbb{C}}^{N\times1}}$  denotes the  received signal vector, $\mathbf{H}\in {{\mathbb{C}}^{N\times K}}$  is the Rayleigh fading channel matrix whose elements are  assumed to be i.i.d.  circularly symmetric complex Gaussian distributed with zero mean and unit variance, $\mathbf{x}\in {\mathcal{O}^{K\times1}}$ represents the transmitted signal vector  whose entries are chosen independently from a  power-normalized constellation $\mathcal{O}$ with  $\left| \mathcal{O} \right| = M$, and
$\mathbf{n}\in {{\mathbb{C}}^{N\times1}}$ models the additive white Gaussian noise vector with zero mean and covariance matrix ${\sigma _{n}^{2}}{\mathbf{I}_{N}}$.
The receiver is assumed to obtain perfect CSI.

\subsection{Decentralized Baseband Processing}

\begin{figure}[!t]
\centering
\includegraphics[width=3.5in]{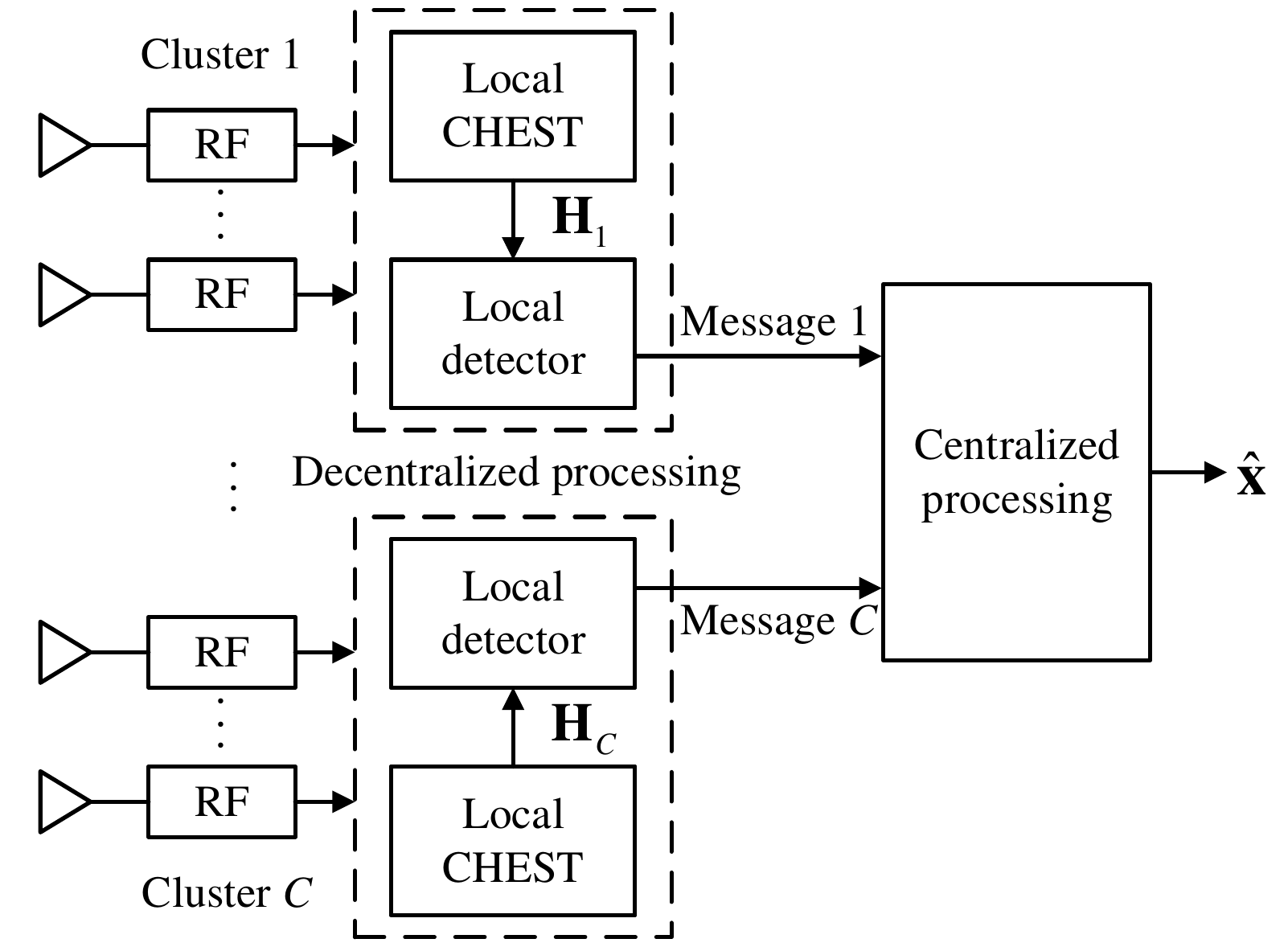}
\caption{Uplink massive MU-MIMO system with DBP architecture.}
\label{fig1}
\end{figure}

In the DBP architecture, the $N$ BS  antennas are partitioned into $C$  independent antenna clusters where $C \in \mathbb N$.
The $c\text{th}$ antenna cluster is equipped with ${{N}_{c}}={{\omega }_{c}}N$ antennas where ${{\omega }_{c}}\in \left[ 0,1 \right]$, $\sum\nolimits_{c=1}^{C}{{{\omega }_{c}}}=1$, and ${{N}_{c}} \in \mathbb N$.
Each antenna cluster contains  local computing hardware, which executes  local RF processing, CHEST, and detection in a decentralized fashion.
A fusion node fuses  the local information passed from each  detector and generates the global estimated symbols.
By partitioning the received signal vector as $\mathbf{y}={{\left[ \mathbf{y}_{1}^{T}, \cdots ,\mathbf{y}_{C}^{T} \right]}^{T}}$,
the local received signal associated with the $c\text{th}$  antenna cluster  is modeled as \cite{dbp-1}
\begin{equation}
\label{equ2}
{{\mathbf{y}}_{c}}={{\mathbf{H}}_{c}}\mathbf{x}+{{\mathbf{n}}_{c}},
\end{equation}
where  the channel matrix is partitioned row-wise into blocks as $\mathbf{H}={{\left[ \mathbf{H}_{1}^{T}, \cdots ,\mathbf{H}_{C}^{T} \right]}^{T}}$, and the noise vector is partitioned as $\mathbf{n}={{\left[ \mathbf{n}_{1}^{T}, \cdots ,\mathbf{n}_{C}^{T} \right]}^{T}}$.
The massive MU-MIMO  with DBP architecture is equivalent to the conventional massive MU-MIMO  when $C=1$.
The block diagram of uplink massive MU-MIMO systems with DBP is depicted in Fig. \ref{fig1}.

\section{Decentralized Gaussian Message Passing  Detection for  Massive MU-MIMO}
\label{sectIII}

\subsection{Decentralized Gaussian Message Passing  Detection}

\begin{figure}[!t]
\centering
\includegraphics[width=3.5in]{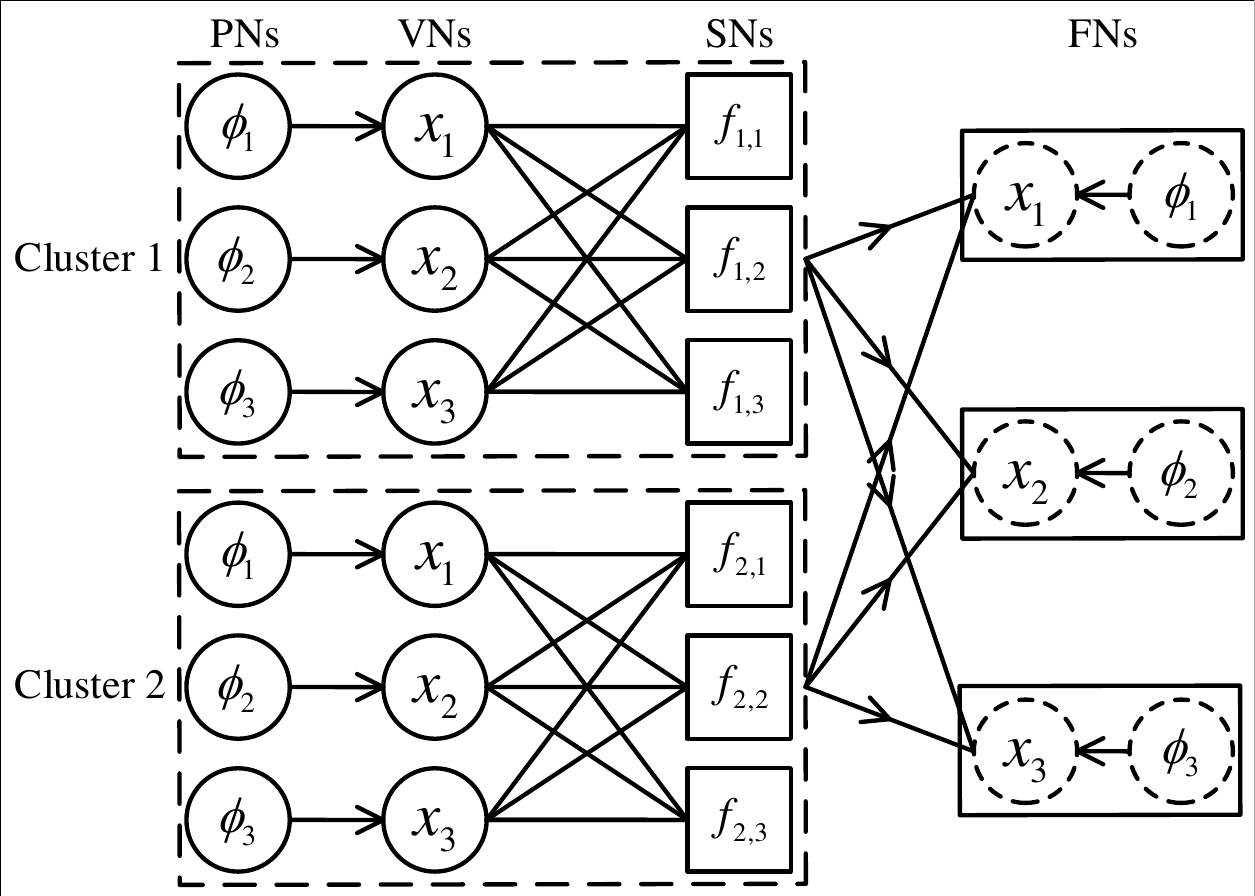}
\caption{Factor graph of decentralized GMP detection for $N\times K=6\times 3$ and $C=2$ with uniform partition.}
\label{fig2}
\end{figure}

The decentralized GMP detection is  operated  on a pairwise factor graph \cite{gmp-4} consisting  of the prior nodes (PNs), variable nodes (VNs), sum nodes (SNs),  and fusion nodes (FNs), which denote  the mapping constraints, users, likelihood functions, and message fusions, respectively.
An  example of a factor graph for decentralized GMP detection is shown  in Fig. \ref{fig2} where $N\times K=6\times 3$ and $C=2$ with uniform  partition.
Let $\mu _{{{x}_{k}}\to {{f}_{c,n}}}^{t}\left( {{x}_{k}} \right)$ and $\mu _{{{f}_{c,n}}\to {{x}_{k}}}^{t}\left( {{x}_{k}} \right)$ denote the messages sent from the $k\text{th}$ ($k=1,\cdots ,K$) VN to the $n\text{th}$ ($n=1,\cdots ,N_c$) SN in the $c\text{th}$ ($c=1,\cdots ,C$) antenna cluster at the $t\text{th}$ ($t=1,\cdots ,T$) iteration and in the opposite direction, respectively, where $T$ is the maximum
number of iterations.
Based on the sum-product algorithm, the message updating rules are given by  \cite{gmp-1,gmp-4}
\begin{equation}
\label{equ3}
\mu _{{{x}_{k}}\to {{f}_{c,n}}}^{t}\left( {{x}_{k}} \right)=\mu _{{{\phi }_{k}}\to {{x}_{k}}}^{}\left( {{x}_{k}} \right)\prod\limits_{{n}'\ne n}{\mu _{{{f}_{c,{n}'}}\to {{x}_{k}}}^{t-1}\left( {{x}_{k}} \right)},
\end{equation}
\begin{equation}
\label{equ4}
\mu _{{{f}_{c,n}}\to {{x}_{k}}}^{t}\left( {{x}_{k}} \right)\!=\!\sum\limits_{\mathbf{x}\backslash {{x}_{k}}}\!{{{f}_{c,n}}\left( \left. {{y}_{c,n}} \right|\mathbf{x} \right)\!\prod\limits_{{k}'\ne k}\!{\mu _{{{x}_{{{k}'}}}\to {{f}_{c,n}}}^{t}\left( {{x}_{{{k}'}}} \right)}},
\end{equation}
where $\mathbf{x}\backslash {x_{k}}$ denotes all the enumerations of $\mathbf{x}$ except for ${x_{k}}$.
The  \emph{a priori} probability  is  given by $\mu _{{{\phi }_{k}}\to {{x}_{k}}}^{t}\left( {{x}_{k}} \right)=1/M$ and the likelihood function  is expressed as
\begin{equation}
\label{equ5}
{{f}_{c,n}}\left( {{y}_{c,n}}|\mathbf{x} \right)\!=\!\frac{1}{\pi {{\sigma _{n}^{2}}}}\!\exp \!\left( -\frac{1}{{{\sigma _{n}^{2}}}}{{\left| {{y}_{c,n}}-\sum\limits_{k}{{{h}_{c,n,k}}{{x}_{k}}} \right|}^{2}} \right),
\end{equation}
where ${{y}_{c,n}}$ denotes the $n\text{th}$ element  of  ${{\mathbf{y}}_{c}}$, and ${{h}_{c,n,k}}$ denotes the complex channel coefficient from  the $k\text{th}$ user to the $n\text{th}$ BS antenna in the $c\text{th}$ antenna cluster. We next concentrate on the decentralized GMP detection in the $c\text{th}$ antenna cluster.

At the VNs, the summation  in (\ref{equ4}) contains a global search over the joint space of  constellation $\mathcal{O}$ of all users except for the $k\text{th}$ user, which results in exponential computational complexity.
To alleviate the computational complexity, ${x_{k}}$ is considered  as a continuous random variable and the message $\mu _{{{f}_{c,n}}\to {{x}_{k}}}^{t}\left( {{x}_{k}} \right)$ is  approximated as a complex Gaussian PDF  $\mathcal{C}\mathcal{N}\left( {{{h}_{c,n,k}}{x}_{k}};m_{{{f}_{c,n}}\to {{x}_{k}}}^{t},v_{{{f}_{c,n}}\to {{x}_{k}}}^{t} \right)$.
According to the product rule of complex Gaussian PDFs \cite{gaussian}, the message passing in (\ref{equ3}) is thus computed as
\begin{equation}
\label{equ6}
\mu _{{{x}_{k}}\to {{f}_{c,n}}}^{t}\!\!\left( {{x}_{k}} \right)\! \approx \!\mu _{{{\phi }_{k}}\to {{x}_{k}}}^{{}}\!\!\left( {{x}_{k}} \right)\!\mathcal{C}\mathcal{N}\!\left( \!{{x}_{k}} ;\!z_{{{x}_{k}}\to {{f}_{c,n}}}^{t-1}\!,\!\gamma _{{{x}_{k}}\to {{f}_{c,n}}}^{t-1} \!\right).
\end{equation}
The variance and mean
are  calculated  as
\begin{equation}
\label{equ7}
\frac{1}{\gamma_{{{x}_{k}}\to {{f}_{c,n}}}^{t-1}}=\sum\limits_{{n}'\ne n}^{{}}{\frac{{{\left| {{h}_{c,{n}',k}} \right|}^{2}}}{v_{{{f}_{c,{n}'}}\to {{x}_{k}}}^{t-1}}},
\end{equation}
\begin{equation}
\label{equ8}
\frac{z_{{{x}_{k}}\to {{f}_{c,n}}}^{t-1}}{\gamma _{{{x}_{k}}\to {{f}_{c,n}}}^{t-1}}=\sum\limits_{{n}'\ne n}^{{}}{\frac{h_{c,{n}',k}^{*}m_{{{f}_{c,{n}'}}\to {{x}_{k}}}^{t-1}}{v_{{{f}_{c,{n}'}}\to {{x}_{k}}}^{t-1}}},
\end{equation}
where $m_{{{f}_{c,n}} \to {{x}_{k}}}^{0}=0$ and $v_{{{f}_{c,n}} \to {{x}_{k}}}^{0}\to +\infty$ are  initialized.
The message $\mu _{{{x}_{k}}\to {{f}_{c,n}}}^{t}\left( {{x}_{k}} \right)$ is also approximated as a complex Gaussian PDF
by minimizing the
KL divergence
${{D}_{KL}}\left(\mu _{{{x}_{k}}\to {{f}_{c,n}}}^{t}\left( {{x}_{k}} \right)|| \mathcal{C}\mathcal{N}\left( {{x}_{k}};m_{{{x}_{k}}\to {{f}_{c,n}}}^{t},v_{{{x}_{k}}\to {{f}_{c,n}}}^{t} \right)\right)$
for arbitrary prior probability.
The mean and variance propagated from VN  to SN  are obtained via moment matching
\begin{equation}
\label{equ9}
m_{{{x}_{k}}\to {{f}_{c,n}}}^{t}=\sum\limits_{\alpha \in \mathcal{O}}{\alpha \mu _{{{x}_{k}}\to {{f}_{c,n}}}^{t}\left( {{x}_{k}}=\alpha  \right)},
\end{equation}
\begin{equation}
\label{equ10}
v_{{{x}_{k}}\to {{f}_{c,n}}}^{t}=\sum\limits_{\alpha \in \mathcal{O}}{{{\left| \alpha  \right|}^{2}}\mu _{{{x}_{k}}\to {{f}_{c,n}}}^{t}\left( {{x}_{k}}=\alpha  \right)}-{{\left| m_{{{x}_{k}}\to {{f}_{c,n}}}^{t} \right|}^{2}}.
\end{equation}

At the SNs, the mean and variance propagated from SN  to VN  are calculated as \cite{gmp-2}
\begin{equation}
\label{equ11}
m_{{{f}_{c,n}}\to {{x}_{k}}}^{t}={{y}_{c,n}}-\sum\limits_{{k}'\ne k}^{{}}{{{h}_{c,n,{k}'}}m_{{{x}_{{{k}'}}}\to {{f}_{c,n}}}^{t}},
\end{equation}
\begin{equation}
\label{equ12}
v_{{{f}_{c,n}}\to {{x}_{k}}}^{t}= \sigma _{n}^{2}+\sum\limits_{{k}'\ne k}^{{}}{{{\left| {{h}_{c,n,{k}'}} \right|}^{2}}v_{{{x}_{{{k}'}}}\to {{f}_{c,n}}}^{t}}.
\end{equation}
The  means and variances are iteratively propagated between VNs and SNs  until $T$ is reached.
When the iterative detection process is terminated, the FNs collect the local
multiplicative messages
$\sum\nolimits_{n}{\mu _{{{f}_{c,n}}\to {{x}_{k}}}^{T}\left( {{x}_{k}} \right)}$, i.e., the local  means $m_{{{f}_{c,n}}\to {{x}_{k}}}^{T}$, and variances $v_{{{f}_{c,n}}\to {{x}_{k}}}^{T}$ from all the antenna clusters to form the global estimation.

At the FNs, the $k\text{th}$  VN gathers information from its input (the $k\text{th}$ PN) and  the SNs of all the antenna clusters to calculate  the global symbol belief $\mu _{{{x}_{k}}}^{{}}\left( {{x}_{k}} \right)$, i.e., the   \emph{a posteriori} probability.
The global symbol belief is given by
\begin{equation}
\label{equ13}
\begin{aligned}
   \mu _{{{x}_{k}}}^{{}}\left( {{x}_{k}} \right)&\propto \mu _{{{\phi }_{k}}\to {{x}_{k}}}^{{}}\left( {{x}_{k}} \right)\prod\limits_{c}{\prod\limits_{n}{\mu _{{{f}_{c,n}}\to {{x}_{k}}}^{T}\left( {{x}_{k}} \right)}} \\
 & \propto\mu _{{{\phi }_{k}}\to {{x}_{k}}}^{{}}\left( {{x}_{k}} \right)\mathcal{C}\mathcal{N}\left( {{x}_{k}};z_{k}^{{}},\gamma _{k}^{{}} \right),
\end{aligned}
\end{equation}
where the variance and mean are  computed as
\begin{equation}
\label{equ14}
\frac{1}{\gamma _{k}^{{}}}=\sum\limits_{c}^{{}}{\sum\limits_{n}^{{}}{\frac{{{\left| {{h}_{c,n,k}} \right|}^{2}}}{v_{{{f}_{c,n}}\to {{x}_{k}}}^{T}}}},
\end{equation}
\begin{equation}
\label{equ15}
\frac{z_{k}^{{}}}{\gamma_{k}^{{}}}=\sum\limits_{c}^{{}}{\sum\limits_{n}^{{}}{\frac{h_{c,n,k}^{*}m_{{{f}_{c,n}}\to {{x}_{k}}}^{T}}{v_{{{f}_{c,n}}\to {{x}_{k}}}^{T}}}}.
\end{equation}
The global symbol belief is  then normalized as
\begin{equation}
\label{equ16}
\mu _{{{x}_{k}}}^{{}}\!\!\left ( {{x}_{k}}\!=\!\alpha  \right)\!=\!\frac{\mu _{{{\phi }_{k}}\to {{x}_{k}}}^{{}}\!\!\left( {{x}_{k}}\!=\!\alpha  \right)\!\mathcal{C}\mathcal{N}\!\left( {{x}_{k}}\!=\!\alpha ;z_{k}^{{}},\gamma _{k}^{{}} \right)}{\sum\limits_{{\alpha }' \in \mathcal{O}}\!\!{\mu _{{{\phi }_{k}}\to {{x}_{k}}}^{{}}\!\!\left( {{x}_{k}}\!=\!{\alpha }'  \right)\!\mathcal{C}\mathcal{N}\!\left( {{x}_{k}}\!=\!{\alpha }' ;z_{k}^{{}},\gamma _{k}^{{}} \right)}},
\end{equation}
and the global  estimated soft symbol is calculated as
\begin{equation}
\label{equ17}
{{\hat{x}}_{k}}=\sum\limits_{\alpha \in \mathcal{O}}^{{}}{\alpha \mu _{{{x}_{k}}}^{{}}\left( {{x}_{k}}=\alpha  \right)}.
\end{equation}
The proposed decentralized GMP  detection for  uplink massive  MU-MIMO systems is summarized in Algorithm \ref{alg1}.


\begin{algorithm}[!t]
\caption{The Proposed Decentralized GMP Detection}
\label{alg1}
\begin{algorithmic}[1]
\\{\bf Input:} ${{\mathbf{y}}_{c}}$, ${{\mathbf{H}}_{c}}$, ${\sigma _{n}^{2}}$, ${{\omega }_{c}}$, $T$
\\{\bf Initialization:} $m_{{{f}_{c,n}}\to {{x}_{k}}}^{0}=0$, $v_{{{f}_{c,n}}\to {{x}_{k}}}^{0}\to +\infty$
\State {\bf Decentralized processing in each antenna cluster:}
\For{$c = 1, \cdots ,C$}
\For{$t = 1, \cdots ,T$}
\State Compute $\mu _{{{x}_{k}}\to {{f}_{c,n}}}^{t}\left( {{x}_{k}} \right)$ for each $k$ and $n$ via (\ref{equ6}), (\ref{equ7}), and (\ref{equ8}).
\State Compute $m_{{{x}_{k}}\to {{f}_{c,n}}}^{t}$ and $v_{{{x}_{k}}\to {{f}_{c,n}}}^{t}$ for each $k$ and $n$ via (\ref{equ9}) and (\ref{equ10}).\
\State Compute $m_{{{f}_{c,n}}\to {{x}_{k}}}^{t}$ and $v_{{{f}_{c,n}}\to {{x}_{k}}}^{t}$ for each $k$ and $n$ via (\ref{equ11})  and (\ref{equ12}).\
\EndFor
\EndFor
\State {\bf Centralized processing based on message fusion rule:}
\For{$k = 1, \cdots ,K$}
\State Compute $\mu _{{{x}_{k}}}^{{}}\left ( {{x}_{k}}=\alpha  \right)$ for each $\alpha \in \mathcal{O}$ via  (\ref{equ13}), (\ref{equ14}), (\ref{equ15}), and (\ref{equ16}).\
\State Compute ${{\hat{x}}_{k}}$ via  (\ref{equ17}).\
\EndFor
\\{\bf Output:} $\mathbf{\hat{x}}=\left[ {{{\hat{x}}}_{1}},{{{\hat{x}}}_{2}},\cdots {{{\hat{x}}}_{K}} \right]$
\end{algorithmic}
\end{algorithm}

\subsection{State Evolution Analysis}

To analyze the state evolution  of  the proposed decentralized GMP detection,
we  concentrate on the large-system limit (i.e., $N \to \infty $,  fixing the system ratio ${K}/{N}$, and  fixing $C$) and Gaussian sources (i.e., $\mathbf{x}\sim \mathcal{C}\mathcal{N}\left( 0,\sigma _{x}^{2}{\mathbf{I}_{K}} \right)$).
With the symmetry  of all the variances \cite{gmp-2,gmp-3} in the $c\text{th}$  antenna cluster, we  assume
$\underset{t\to +\infty }{\mathop{\lim }}\,v_{{{x}_{k}}\to {{f}_{c,n}}}^{t}\!=\!{v_{x\to f}^{c}}$,   $\underset{t\to +\infty }{\mathop{\lim }}\,v_{{{f}_{c,n}}\to {{x}_{k}}}^{t}\!=\!{v_{f\to x}^{c}}$, and
$\mathbb{E}\left[ {{\left| {{h}_{c,n,k}} \right|}^{2}} \right]\!\approx \!1$
for
$\forall k\in \left\{ 1,\cdots ,K \right\}$ and $\forall n\in \left\{ 1,\cdots ,N_c \right\}$ in the large-system limit.
When Gaussian sources are assumed, the variance $v_{{{x}_{k}}\to {{f}_{c,n}}}^{t}$ of  $\mu _{{{x}_{k}}\to {{f}_{c,n}}}^{t}\left( {{x}_{k}} \right)$ (\ref{equ6}) is computed as
\begin{equation}
\label{equ18}
\frac{1}{v_{{{x}_{k}}\to {{f}_{c,n}}}^{t}}=\frac{1}{\sigma _{x}^{2}}+\sum\limits_{{n}'\ne n}^{{}}{\frac{{{\left| {{h}_{c,{n}',k}} \right|}^{2}}}{v_{{{f}_{c,{n}'}}\to {{x}_{k}}}^{t}}}.
\end{equation}
When $t \to  +\infty$,
taking the expectations of (\ref{equ12}) and (\ref{equ18})
results in \cite{gmp-2}
\begin{equation}
\label{equ19}
v_{f\to x}^{c}=\sigma _{n}^{2}+Kv_{x\to f}^{c},
\end{equation}
\begin{equation}
\label{equ20}
\frac{1}{v_{x\to f}^{c}}=\frac{1}{\sigma _{x}^{2}}+\frac{{{\omega }_{c}}N}{v_{f\to x}^{c}}.
\end{equation}
Combining (\ref{equ19}) and (\ref{equ20}), we have the following equation
\begin{equation}
\label{equ21}
v{{_{f\to x}^{c}}^{2}}+\left( {{\omega }_{c}}N\sigma _{x}^{2}-K\sigma _{x}^{2}-\sigma _{n}^{2} \right)v_{f\to x}^{c}-{{\omega }_{c}}N\sigma _{x}^{2}\sigma _{n}^{2}=0.
\end{equation}
The fix point is computed as the positive solution
\begin{equation}
\label{equ22}
v_{\!f\!\to  x}^{c}\!\!=\!\!\frac{K\!\sigma _{\!x}^{2}\!-\!{{\omega }_{\! c}} N\!\sigma _{\!x}^{2}\!+\!\sigma _{\!n}^{2}\!+\!\!\sqrt{\!{{\left( {{\omega }_{\! c}} N\!\sigma _{\!x}^{2}\!-\!K\!\sigma _{\!x}^{2}\!+\!\sigma _{\!n}^{2} \right)}^{2}}\!\!+\!4 K\!\sigma _{\!x}^{2}\sigma _{\!n}^{2}}}{2}.
\end{equation}
When considering Gaussian sources, the global  symbol belief
is also a Gaussian function whose variance $v$ is calculated as
\begin{equation}
\label{equ23}
\begin{aligned}
  &  \frac{1}{v }  =\frac{1}{\sigma _{x}^{2}}+\sum\limits_{c}^{{}}{\frac{{{\omega }_{c}}N}{v_{f\to x}^{c}}}=\frac{1}{\sigma _{x}^{2}}+ \\
 & \!\sum\limits_{c}^{{}}\!\!{\frac{2{{\omega }_{c}}N}{K\!\sigma _{\!x}^{2}\!-\!{{\omega }_{\! c}} N\!\sigma _{\!x}^{2}\!+\!\sigma _{\!n}^{2}\!+\!\!\sqrt{\!{{\left( {{\omega }_{\! c}} N\!\sigma _{\!x}^{2}\!-\!K\!\sigma _{\!x}^{2}\!+\!\sigma _{\!n}^{2} \right)}^{2}}\!\!+\!4 K\!\sigma _{\!x}^{2}\sigma _{\!n}^{2}}}}.
\end{aligned}
\end{equation}

\subsection{Partition Scheme of the Antenna Clusters}

The variance of the symbol belief affects the   convergence rate of the proposed decentralized GMP detection.
A smaller variance results in a larger  probability  for the most probable  constellation symbol, which accelerates the convergence rate of the symbol belief.
We define the function $f\left( x\right)$ as
\begin{equation}
\label{equ24}
f\!\left( x\right)\!=\!\!\frac{2{x}\!N}{K\!\sigma _{\!x}^{2}\!-\!x\! N\!\sigma _{\!x}^{2}\!+\!\sigma _{\!n}^{2}\!+\!\!\sqrt{\!{{\left( x\! N\!\sigma _{\!x}^{2}\!-\!K\!\sigma _{\!x}^{2}\!+\!\sigma _{\!n}^{2} \right)}^{2}}\!\!+\!4 K\!\sigma _{\!x}^{2}\sigma _{\!n}^{2}}},
\end{equation}
where $x\in \left[ 0,1 \right]$.
Note that $f\left( x  \right)$ is convex in $x$ as the second order derivative
${f}''\left( x \right)> 0$.
According to the Jensen inequality, we obtain that
\begin{equation}
\label{equ25}
\frac{1}{C}\sum\limits_{c=1}^{C}{f\left( {{\omega }_{c}} \right)}\ge f\left( \frac{1}{C}\sum\limits_{c=1}^{C}{{{\omega }_{c}}} \right),
\end{equation}
where  equality holds if ${{\omega }_{1}}={{\omega }_{2}}=\cdots ={{\omega }_{C}}$.
This indicates that the uniform partition results in
the slowest convergence rate of the symbol belief.
From (\ref{equ25}) we obtain  $\sum\nolimits_{c=1}^{C}{f\left( {{\omega }_{c}} \right)}\ge Cf\left( {1}/{C}\; \right)$.
We define the function $g\left( x \right)={f\left( x \right)}/{x}$.
Note that $g\left( x \right)$ is monotonically increased with the increase of $x$ as the first order derivative ${g}'\left( x \right)> 0$.
This indicates that a smaller number of antenna clusters results in a faster convergence rate of the symbol belief when using the uniform partition.

\subsection{Computational Complexity Analysis}

The computational complexity is analyzed  in terms of the required number of complex multiplications.
The complexity of the decentralized processing is $NK+8TNK+6MTNK$ while the centralized processing requires $5MK$ complex multiplications.
The computational complexity of the proposed decentralized GMP detection  and the recently proposed decentralized algorithms is summarized in  Table \ref{tab1}.

\begin{table}[!t]
\renewcommand{\arraystretch}{1.8}
\caption{Computational Complexity.}
\label{tab1}
\centering
\begin{tabular}{c || c}
\hline
\hline
Algorithms & Number of complex multiplications\\
\hline
GMP & $\left( 6MTN+8TN+N+5M \right)K$\\
\hline
LAMA & $\left( N+CT \right){{K}^{2}}+\left( N+5CMT+2CT+3C \right)K$\\
\hline
MMSE  & $\frac{10}{3}C{{K}^{3}}+\left( N+\frac{7}{2}C \right){{K}^{2}}+\left( N+\frac{13}{6}C \right)K$\\
\hline
MRC  & $C{{K}^{3}}+\left( N+3C \right){{K}^{2}}+\left( N+5C \right)K$\\
\hline
\hline
\end{tabular}
\end{table}

\begin{figure}[!t]
\centering
\includegraphics[width=3.5in]{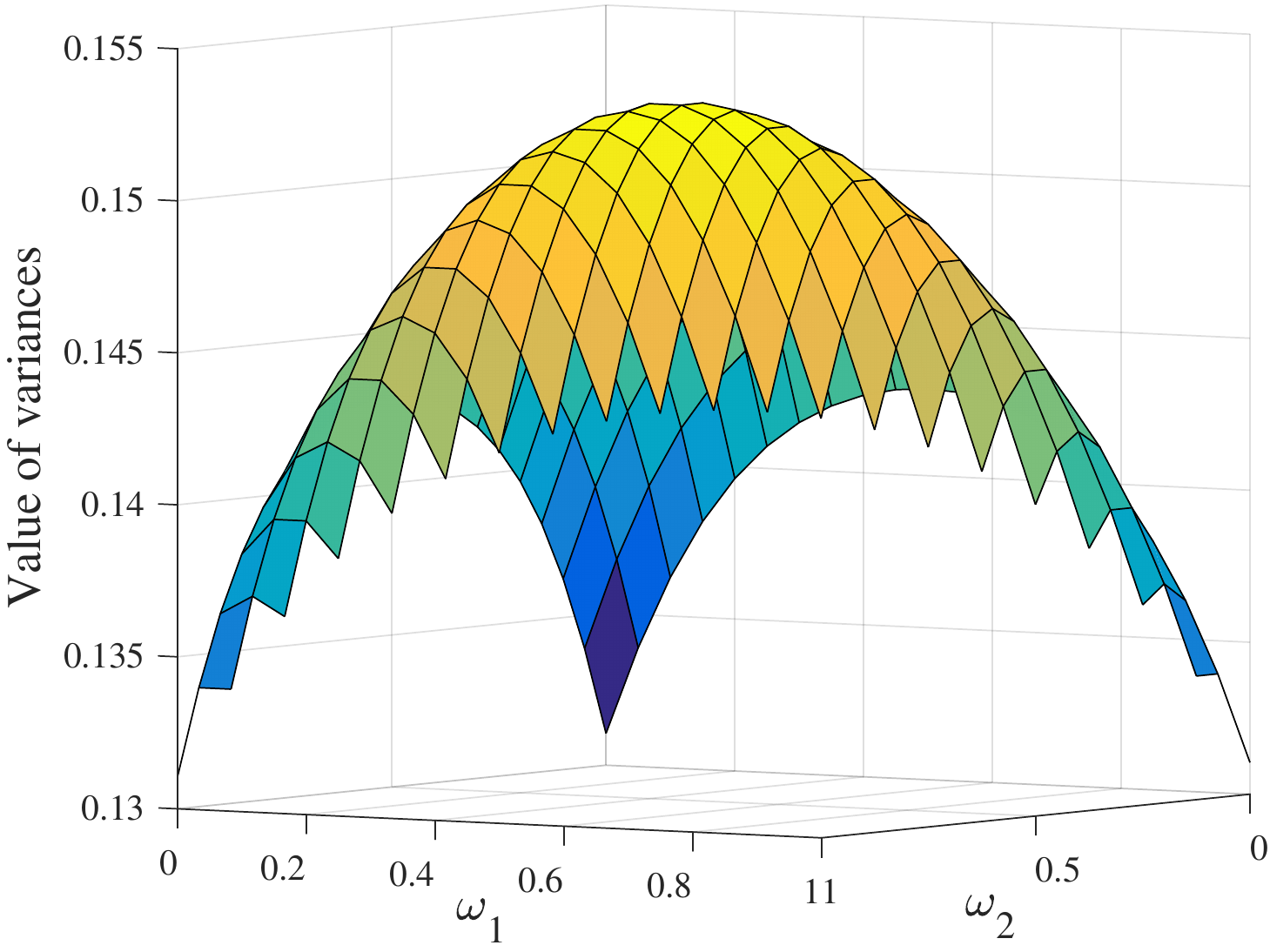}
\caption{Variance  variation  of symbol beliefs versus $\omega$ at $\text{SNR}=0\thinspace\text{dB}$ where $K=16$ and $C=3$ with uniform partition.}
\label{fig3}
\end{figure}

\begin{figure}[!t]
\centering
\includegraphics[width=3.5in]{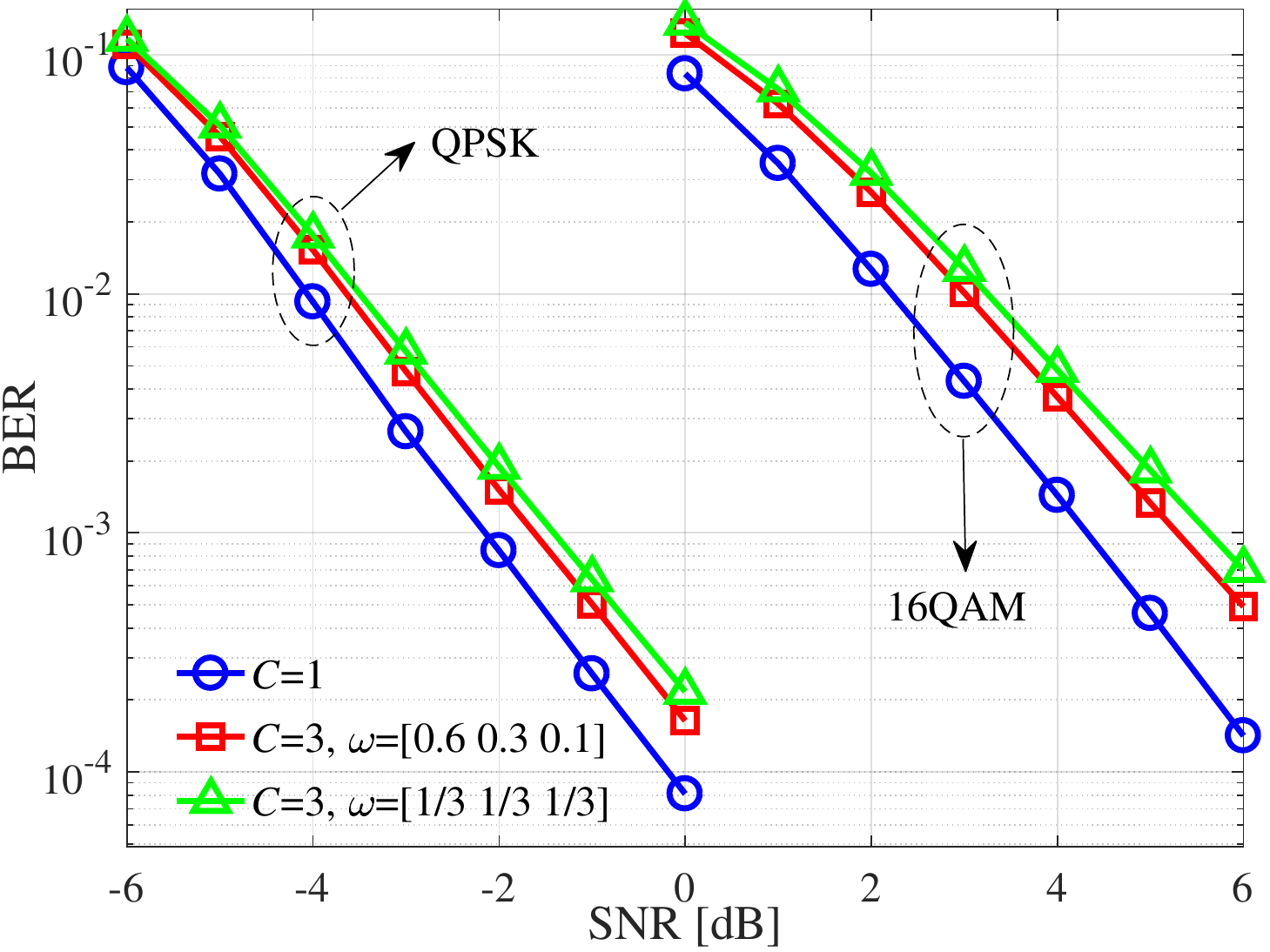}
\caption{BER performance of the decentralized GMP detection for different antenna cluster  partition schemes where $K=16$.}
\label{fig4}
\end{figure}

\section{Simulation Results}
\label{sectIV}

In this section, we evaluate the BER performance and computational complexity of the proposed decentralized GMP detection.
The recently proposed decentralized LAMA, MMSE, and MRC  algorithms are compared as benchmarks.
The BS is equipped with $N=120$ antennas.
The  convolutional code of rate $1/2$ is adopted.
The maximum number of iterations for the decentralized GMP and LAMA methods is set as $T=5$.


\begin{figure}[!t]
\centering
\includegraphics[width=3.5in]{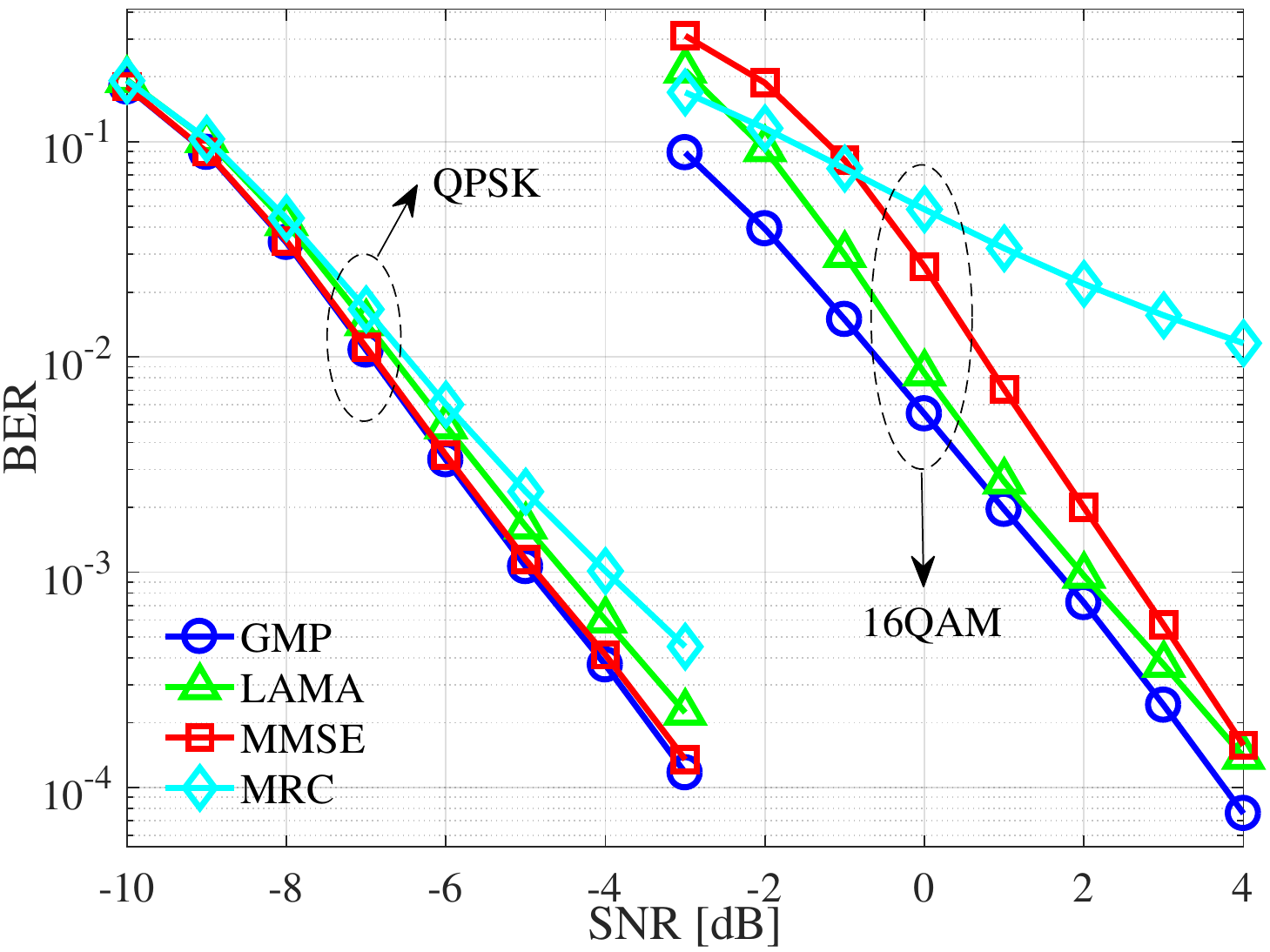}
\caption{BER performance of detection algorithms  where $K=8$ and $C=3$ with uniform partition.}
\label{fig5}
\end{figure}

\begin{figure}[!t]
\centering
\includegraphics[width=3.5in]{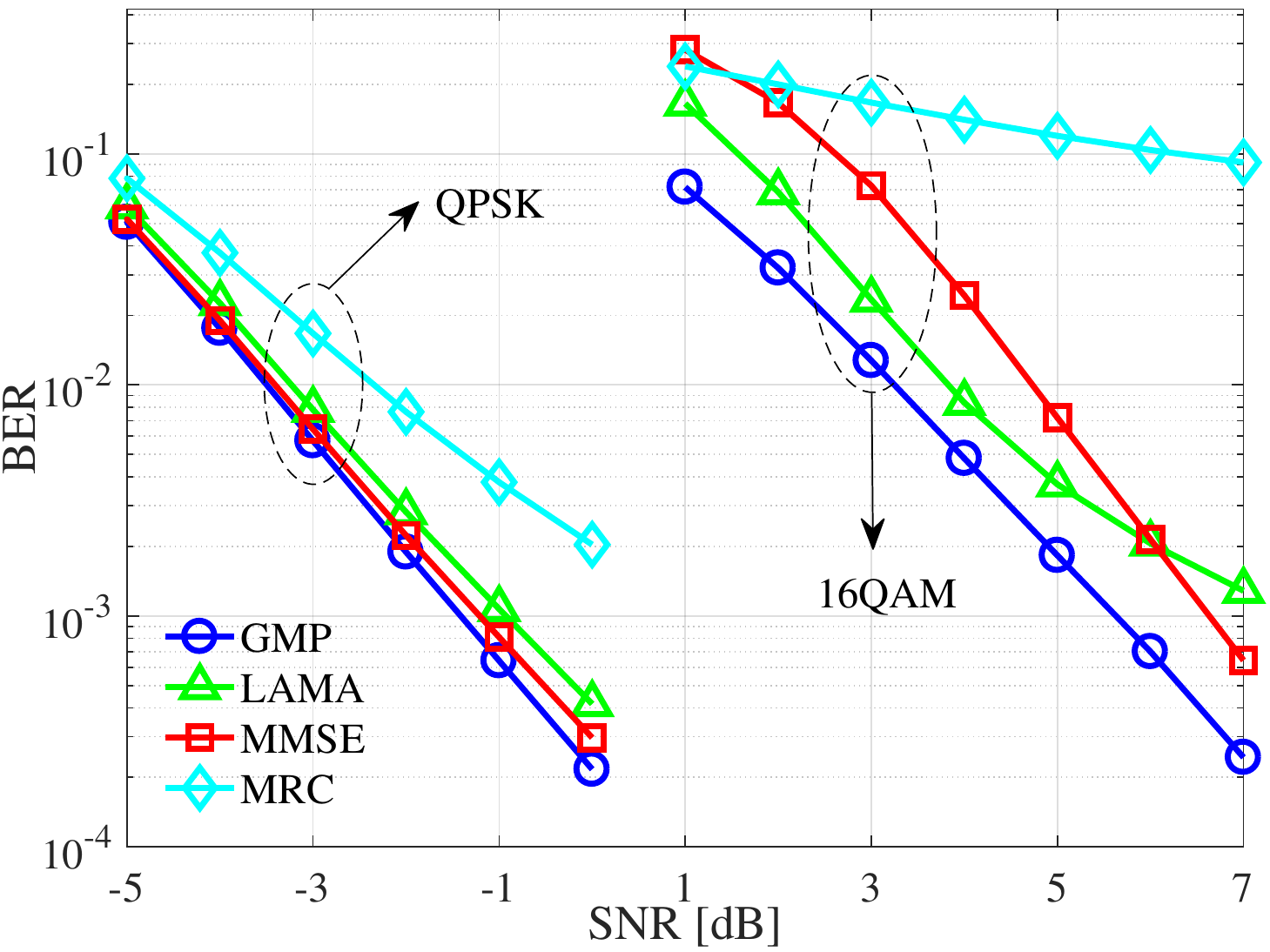}
\caption{BER performance of detection algorithms  where $K=16$ and $C=3$ with uniform partition.}
\label{fig6}
\end{figure}

Fig. \ref{fig3} shows the variance variation of symbol beliefs versus $\omega$ at $\text{SNR}=0\thinspace\text{dB}$, where $K = 16$ and $C = 3$ with uniform partition. This figure further illustrates that the uniform partition leads to the largest variances.
Fig. \ref{fig4} evaluates the BER performance of the proposed decentralized GMP detection with different  antenna cluster partition schemes for $C=3$.
The decentralized processing with nonuniform antenna cluster partition outperforms the  uniform counterpart and  approaches the centralized processing with acceptable  performance loss.

To prove the efficiency of the proposed decentralized GMP detection, the BER performance comparison with the state-of-the-art decentralized algorithms is evaluated in equally-sized antenna clusters which are desirable in practice as the uniform partition minimizes the interconnect or chip  input/output  bandwidth as well as the computational complexity per computing fabric \cite{dbp-1}.
Fig. \ref{fig5} and Fig. \ref{fig6} illustrate that the proposed decentralized GMP detection outperforms the other  methods, especially for high  modulation order and large   number of user.
For example, the decentralized GMP achieves gains of nearly $1.1\thinspace\text{dB}$ and $2.1\thinspace\text{dB}$ over the decentralized MMSE and LAMA at BER of ${{10}^{-3}}$, respectively, for $K=16$ and $16$QAM.

\begin{figure}[!t]
\centering
\includegraphics[width=3.5in]{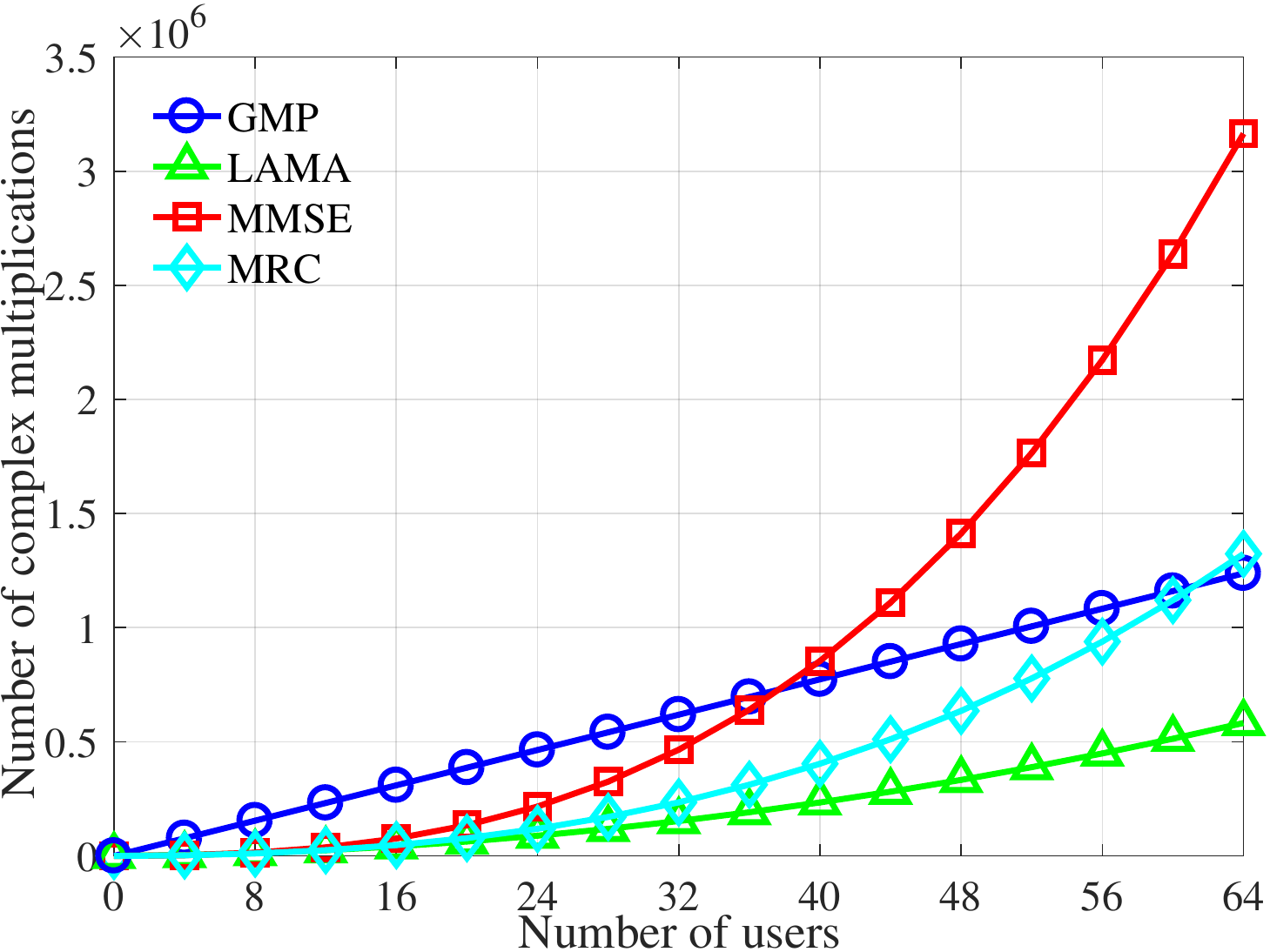}
\caption{Computational complexity comparison  where  $C = 3$ and QPSK.}
\label{fig7}
\end{figure}

Fig. \ref{fig7} presents the complexity comparison which shows that the proposed  decentralized GMP detection behaves linear  complexity with the increase of user number. This indicates that the implementation of the proposed algorithm exhibits  hardware-friendly complexity in the case of  massive connection.

\section{Conclusion}
\label{sectV}

This paper proposed a  novel decentralized  GMP detection for uplink massive MU-MIMO systems with DBP.
The centralized processing executes the message fusion   by gathering the local messages propagated from each antenna cluster  in parallel.
The convergence of the proposed algorithm is characterized by state evolution under the assumptions of large-system limit and Gaussian sources.
We demonstrated that the nonuniform partition outperforms  the  uniform partition for  a fixed antenna cluster number.
Simulation results showed that the proposed decentralized GMP detection outperforms the recently proposed methods and exhibits linear  complexity.

\ifCLASSOPTIONcaptionsoff
  \newpage
\fi


\begin{thebibliography}{99}


\bibitem{m-mimo-1}
T. L. Marzetta, ``Noncooperative Cellular Wireless with Unlimited Numbers of Base Station Antennas,"  \emph{IEEE Trans. Wireless Commun.}, vol. 9, no. 11, pp. 3590-3600, November 2010.




\bibitem{dbp-1}
C. Jeon, K. Li, J. R. Cavallaro, and C. Studer, ``Decentralized Equalization With Feedforward Architectures for Massive MU-MIMO," \emph{IEEE Trans. Signal Process.}, vol. 67, no. 17, pp. 4418-4432, 2019.

\bibitem{dbp-2}
K. Li, R. R. Sharan, Y. Chen, T. Goldstein, J. R. Cavallaro, and C. Studer, ``Decentralized Baseband Processing for Massive MU-MIMO Systems,"  \emph{IEEE J. Emerg. Sel. Topics Circuits Syst.}, vol. 7, no. 4, pp. 491-507, Dec. 2017.


\bibitem{dbp-3}
K. Li, Y. Chen, R. Sharan, T. Goldstein, J. R. Cavallaro, and C. Studer, ``Decentralized data detection for massive MU-MIMO on a Xeon Phi cluster,” in \emph{Proc. 50th Asilomar Conf. Signals, Syst. Comput.}, Pacific
Grove, CA, 2016, pp. 468-472.

\bibitem{dbp-4}
K. Li, O. Castaneda, C. Jeon, J. R. Cavallaro, and C. Studer, ``Decentralized coordinate-descent data detection and precoding for massive MU-MIMO,” in \emph{Proc. IEEE Int. Symp. Circuits Syst.}, Sapporo, Japan, 2019, pp. 1-5.

\bibitem{dbp-5}
C. Jeon, K. Li, J. R. Cavallaro, and C. Studer, ``On the achievable rates of decentralized equalization in massive MU-MIMO systems,” in \emph{Proc. IEEE Int. Symp. Inf. Theory}, Aachen, 2017, pp. 1102-1106.









\bibitem{gmp-1}
S. Wu, L. Kuang, Z. Ni, J. Lu, D. Huang, and Q. Guo, ``Low-Complexity Iterative Detection for Large-Scale Multiuser MIMO-OFDM Systems Using Approximate Message Passing,"  \emph{IEEE J. Sel. Topics Signal Process.}, vol. 8, no. 5, pp. 902-915, Oct. 2014.



\bibitem{gmp-2}
L. Liu, C. Yuen, Y. L. Guan, Y. Li, and Y. Su, ``Convergence Analysis and Assurance for Gaussian Message Passing Iterative Detector in Massive MU-MIMO Systems," \emph{IEEE Trans. Wireless Commun.}, vol. 15, no. 9, pp. 6487-6501, 2016.

\bibitem{gmp-3}
L. Liu, C. Yuen, Y. L. Guan, Y. Li, and C. Huang, ``Gaussian Message Passing for Overloaded Massive MIMO-NOMA," \emph{IEEE Trans. Wireless Commun.}, vol. 18, no. 1, pp. 210-226, 2019.




\bibitem{gmp-4}
F. R. Kschischang, B. J. Frey, and H.-A. Loeliger, ``Factor graphs and the sum-product algorithm,”  \emph{IEEE Trans. Inf. Theory}, vol. 47, no. 2, pp. 498-519, Feb. 2001.


\bibitem{gaussian}
P. Bromiley, ``Products and convolutions of Gaussian probability density
functions,” Tina-Vision Memo Tech. Rep. 2003-003, 2003.


\end{thebibliography}
\end{document}